\documentclass{llncs}
\usepackage{fullpage}
\usepackage{amssymb}
\usepackage{amsmath}
\usepackage{graphicx}
\usepackage{subfigure}
\begin{document}
\mainmatter              
\title{On the Structure of Equilibria in Basic Network Formation}
%
%
\author{S. Nikoletseas\inst{3,4}\thanks{Research partially supported by MULTIPLEX project - 317532.} \and P. Panagopoulou\inst{4} \and C. Raptopoulos\inst{1,4}\thanks{Research supported by SHARPEN project - PE6 (1081). The research project is implemented within the framework of the Action ``Supporting Postdoctoral Researchers'' of the Operational Program ``Education and Lifelong Learning'' (Action's Beneficiary: General Secretariat for Research and Technology), and is co-financed by the European Social Fund (ESF) and the Greek State.} \and P. G. Spirakis \inst{2,3,4}\thanks{Research partially supported by MULTIPLEX project - 317532.}}
%
%
%
\institute{Computer Science Department, University of Geneva, Switzerland \\
\and Computer Science Department, University of Liverpool, United Kingdom \\
\and Computer Science Department, University of Patras, Greece\\
\and Computer Technology Institute and Press ``Diophantus'', Greece \\
\email{nikole@cti.gr}, \email{panagopp@cti.gr}, \email{raptopox@ceid.upatras.gr}, \email{spirakis@cti.gr}}
\maketitle              

\begin{abstract}
We study network connection games where the nodes of a network perform edge swaps in order to improve their communication costs. For the model proposed by \cite{ADHL10}, in which the selfish cost of a node is the sum of all shortest path distances to the other nodes, we use the probabilistic method to provide a new, structural characterization of equilibrium graphs. We show how to use this characterization in order to prove upper bounds on the diameter of equilibrium graphs in terms of the size of the largest $k$-vicinity (defined as the the set of vertices within distance $k$ from a vertex), for any $k \geq 1$ and in terms of the number of edges, thus settling positively a conjecture of \cite{ADHL10} in the cases of graphs of large $k$-vicinity size (including graphs of large maximum degree) and of graphs which are dense enough. 

Next, we present a new swap-based network creation game, in which selfish costs depend on the immediate neighborhood of each node; in particular, the profit of a node is defined as the sum of the degrees of its neighbors. We prove that, in contrast to the previous model, this network creation game admits an exact potential, and also that any equilibrium graph contains an induced star. The existence of the potential function is exploited in order to show that an equilibrium can be reached in expected polynomial time even in the case where nodes can only acquire limited knowledge concerning non-neighboring nodes. 
\end{abstract}
\section{Introduction}

Large scale networks, such as the Internet, are built, maintained and used by selfish entities, all of whom aim at optimizing their own cost and quality of network usage. This suggests that such individual entities will have an incentive to form connections with others to shape the network in ways that are advantageous to themselves. A \emph{network creation game} specifies a set of players, the link formation actions available to each player and the payoffs to each player from the networks that arise out of the link formation action profiles adopted by the players. There are many potential models of network creation games that can be developed, depending on which individuals have the decision power to form or delete a link (i.e., which subset of nodes forms the set of players, and what are their available strategies), as well as on the specification of the payoff allocation rule (i.e., what is the cost of forming or deleting a link and under which objectives do the selfish individuals evaluate the quality of a network).

The study of network creation games focuses on \emph{strategically stable} or \emph{equilibrium} networks, i.e., networks where there are no incentives for individual players to form or delete links and thereby alter the network. We are interested in both \emph{static} and \emph{dynamic} properties of equilibrium networks. Static properties include the structure (i.e., the topologies) of equilibrium networks, as well as bounds on the \emph{Price of Anarchy} \cite{KP99} and the \emph{Price of Stability} \cite{ADK+04} in network creation games, which, roughly speaking, measure the effect of selfish link formation on the quality of the resulting network. On the other hand, dynamic properties of equilibrium networks specify whether (and how fast) selfish players can actually converge into a stable network and thus find a desired equilibrium.

Fabrikant et al. \cite{FLMPS03} introduced a simple network formation game, where each player is identified with a node, and each node can choose to create a link between itself and any subset of other nodes. Each link requires a fixed cost $\alpha>0$ to be built, and each player has two competing goals: to pay for as few links as possible, and to minimize the distance to all other players in the resulting network. In particular, the objective of each player is to minimize the sum of costs of the links created by itself plus the sum of distances to all other nodes of the resulting network. In such a setting, the network is in equilibrium if no node can improve her objective cost by deleting and/or creating \emph{any subset} of incident links. Fabrikant et al. \cite{FLMPS03} proved an upper bound of $O(\alpha)$ on the Price of Anarchy. Subsequent work \cite{AEE+06,DHM+07,MS10} showed that the Price of Anarchy is constant for almost all values of $\alpha$. 

This model of a network creation game is simple enough, while it achieves to capture the flexibility of nodes to create and delete links as well as the trade-off between the cost of creating links and the cost of reaching the other nodes of the network. However, it has a main drawback: it is NP-hard to compute a \emph{best response} of a node. That is, a (computationally bounded) node can not decide which links to add and which to remove in order to achieve a minimal cost, given the strategies of the other nodes. This implies that the players of the game can not even decide whether they are in equilibrium or not, which further implies that the nodes are incapable of converging into an equilibrium network by performing local selfish improvements. 

In view of the above drawback, Alon et al. \cite{ADHL10} proposed a simpler model, namely the \emph{basic network creation game}. In this game the nodes are significantly less flexible in creating and deleting links: in particular, a node can only \emph{swap} an existing link with another, i.e., delete an incident link and create a new incident link. Alon et al. \cite{ADHL10} considered two different objectives for the nodes, yielding two versions of a basic network creation game: In the sum version, the cost of a node is the sum of its distances to all other nodes, while in the max version the cost of a node is the greatest distance between itself and any other node (i.e., the \emph{eccentricity} of the node, using graph-theoretic terminology). A network is in \emph{swap equilibrium} if no node can decrease its cost by deleting an incident link and creating a new one. With this restriction on the available strategies of each node, it is easy to see that swap equilibria (under either cost objective) can be detected in polynomial time: each node simply has to check each possible swap of a non-neighboring node with a neighboring one. 

Alon et al. \cite{ADHL10} focused on the structure of equilibrium networks of basic network creation games and, in particular, on bounding the diameter of equilibrium networks. For the sum version they gave an upper bound of $2^{O(\sqrt{\ln n})}$ ($n$ being the number of nodes), a lower bound of 3, and a tight bound of 2 for trees. For the diameter version they gave a lower bound of $\Omega(\sqrt{n})$ and a tight upper bound of 3 for trees. Recently, Lenzner \cite{L11} studied the best response dynamics of the sum version of basic network creation games. He proved that, when played on a tree, the game admits an ordinal potential function \cite{MS96}, implying that any selfish improvement sequence is guaranteed to converge to an equilibrium tree. In addition, it was shown that a cubic upper bound of selfish steps is needed in order to reach an equilibrium tree. For general networks however (i.e., networks with cycles), it was shown that the game allows best-response cycles, implying that the game does no longer admit any kind of potential function, which further implies that selfish improvements do not necessarily converge to an equilibrium network.

In a recent paper \cite{MS12} the authors study the concept of asymmetric swap equilibrium and show how this concept generalizes and unifies some other equilibrium concepts for network creation games (as the network creation game of \cite{FLMPS03} and the bounded-budget network creation game of \cite{EFMSSSS11}). Their main result has some similarities to part A in our Lemma \ref{SSEgraphstructure}, but we chose to present our proof here for completeness and because it is different and shorter. The authors in \cite{MS12} use their results to settle the conjecture of \cite{ADHL10} in the case where the minimum degree of an induced subgraph of the equilibrium graph is at least $n^\epsilon$, for some $\epsilon>0$. In contrast, our results do not require any assumptions on the minimum degree and equilibrium graphs may even have vertices of degree 1.

\subsection{Our Contribution}

This paper has two main contributions which are briefly described here. First, for the sum version of the model of \cite{ADHL10}, we provide in Theorem \ref{maintheorem} a new, structural characterization of equilibrium graphs, which roughly states that for any two vertices of degree greater than 1, the majority of the rest of the vertices are almost equidistant from them. The proof uses the probabilistic method, combined with some basic properties of equilibrium graphs. Our characterization can also be seen as a stronger ``skewness'' property like the one defined by the authors in \cite{ADHL10} (see Section 5 in that paper). In fact, using Theorem 1, we show how we can prove upper bounds on the diameter of equilibrium graphs in terms of the size of the largest $k$-vicinity, for any $k \geq 1$ and in terms of the number of edges. As shown in Corollary \ref{maxdegreecorollary} and in Theorem \ref{densecase}, this partially settles positively a conjecture of \cite{ADHL10} (that equilibria graphs have poly-logarithmic diameter), in the cases (a) of graphs that have a vertex with large $k$-vicinity (including graphs with sufficiently large maximum degree) and (b) of graphs which are dense enough.

Even though the model of \cite{ADHL10} is more basic than the model of \cite{FLMPS03} (in that better responses can be determined in polynomial time), it still relies on the fact that each vertex/player has global knowledge of the graph, which is needed to compute its cost function. Furthermore, it is so far unknown how an equilibrium can be reached in a distributed, uncoordinated manner, starting from any initial graph configuration. In many cases though, especially for large-scale networks like the Internet, there is no coordination between nodes and only local information is at their immediate disposal. To address these issues, we present, as our second contribution in this paper, a new swap-based network creation game, in which selfish costs depend on the immediate neighborhood of each vertex/player. In particular, for each vertex, we define its profit to be the sum of the degrees of its neighbors, which is also related to the number of paths of length 2 from that vertex. We prove that, unlike the model of \cite{ADHL10}, this network creation game admits an exact potential, and also that any equilibrium graph contains an induced star. The existence of a potential function implies that better response dynamics always converge to an equilibrium graph within a polynomial number of steps in the number of vertices. Furthermore, we consider a case where vertices can only acquire limited knowledge concerning non-neighboring vertices and we show that we can reach equilibrium in expected polynomial time.

\subsection{Organization of the Paper}

The model of Alon et al. \cite{ADHL10} together with some first useful results are presented in Section \ref{modelandfirst}. In Section \ref{structuralcharacterization} we present the proof of our main theorem, characterizing graphs in Sum-Swap Equilibrium (SSE). Furthermore, we provide some consequences of our characterization in the cases of dense and large degree graphs in Section \ref{largeanddense}. Section \ref{ourmodel} is devoted to the definition of our new model of network formation with local costs. The proof that our network formation game admits a potential is given in Section \ref{potentialproof}. In Section \ref{lowqueries} we consider the case where vertices can have limited knowledge of the graph. Finally, in Section \ref{conclusions}, we provide some concluding remarks and open questions.


\subsection{Notation}

Let $G = (V, E)$ be an undirected graph. For a vertex $v \in V$, we denote by $N_G(v)$ the set of neighbors of $v$ in $G$. We will denote by $\Delta_G$ the maximum degree of a vertex of $G$, i.e., $\Delta_G = \max_{v\in V} \deg(v)$, where $\deg(v)=|N_G(v)|$. For a vertex $v$ of a graph $G$, we will denote by $\deg_{-1}(v)$ the number of neighbors of $v$ that have degree at least 2. We will also denote by $N_{-1}(v)$ the set of neighbors of $v$ having degree at least 2, i.e. $\deg_{-1}(v) = |N_{-1}(v)|$.

For any two vertices $v, u \in V$ we will denote by $dist_G(u, v) = dist_G(v, u)$ the length of a shortest path between $u$ and $v$ in $G$. We denote by $diam(G)$ the diameter of $G$, defined as $diam(G) = \max_{u \in V, v \in V} dist_G(u,v)$. For any vertex $v \in V$ we will denote by $W_{G}(v)$ the sum of distances of all vertices from $v$ in $G$, i.e., $W_G(v) = \sum_{u \in V} dist_G(u, v)$. If the graph is disconnected, then we define $W_{G}(v)$ to be infinite.  

For any subset of vertices $S \subseteq V$, we denote by $G[S]$ the subgraph of $G$ induced on the set $S$. Finally, following the notation in \cite{ADHL10}, for any vertex $u \in V$ and any $k$, we will denote by $B_u(k)$ the $k$-vicinity of $u$, i.e. $B_u(k) \stackrel{def}{=} \{w : dist(w, u) \leq k\}$. 
We will sometimes omit the subscripts $G$ in the above notation if the graph is understood from the context.

\section{Diameter in the Sum Version of Swap Equilibria} \label{Alonmodel}

In this section we consider the sum version of the model of Alon et. al \cite{ADHL10}. In particular, in Theorem \ref{maintheorem} we provide a structural characterization of graphs $G$ that are in Sum-Swap Equilibrium (SSE). Using this characterization we provide an upper bound on the diameter of any swap equilibrium graph which depends on the size of its largest $k$-vicinity, for any $k >0$. 

\subsection{The Model and some First Results} \label{modelandfirst}
We will first describe more formally the model of a basic network creation game proposed in \cite{ADHL10}, and which we consider in this section. We are given an undirected  graph $G=(V,E)$, where each vertex corresponds to a player. The \emph{connection cost} of player $v \in V$ is the sum of distances between $v$ and all other vertices, i.e., equals $W_G(v)$. A player can perform ``edge swaps'', i.e., replace an incident existing edge with another incident edge. More formally, let $u \in N_G(v)$ and $w \notin N_G(v)$. Then, the edge swap $(u,w)$ of $v$ removes the edge $\{v,u\}$ and creates the edge $\{v,w\}$. Therefore, the set of pure strategies of player $v \in V$ in graph $G$ is $S_G(v) = \left\{ N_G(v) \times \{V \setminus \{N_G(v)\cup \{v\}\}\}\right\}$. Observe that the set of pure strategies of a player depends on the current graph $G$ and that an edge swap performed by a player modifies the graph. 

We say that a graph is in \emph{swap sum-equilibrium} (SSE in short) if no player (vertex) can improve her connection cost by performing an edge swap. More formally, we denote by $G_{s_v}$ the graph obtained from $G$ when player $v$ performs the edge swap $s_v \in S_G(v)$. Then, we say that the graph $G=(V,E)$ is in SSE if for all $v \in V$, $W_G(v) \leq W_{G_{s_v}}(v)$ for all $s_v \in S_G(v)$.

We now prove some basic structural properties of graphs in SSE. To avoid trivialities, all graphs considered here will be connected. 

\begin{lemma} \label{SSEgraphstructure}
Let $G$ be a graph in swap sum-equilibrium and let $n$ be the number of vertices. Then the following hold:

\begin{description}

\item[A.] Let $u, v$ be any two vertices of degree greater than 1. Then, there are at least two paths from $u$ to $v$ such that the first edge on each path is different.

\item[B.] If $G$ has a vertex of degree 2, then $diam(G) \leq 9$.





\end{description}
\end{lemma}
\proof \emph{Part (A.):} For some integer $k \geq 1$, let $P$ be the $u-v$ path $u=v_0, v_1, \ldots, v_k=v$. Assume for the sake of contradiction that all $u-v$ paths use the edge $\{u, v_1\}$ (i.e., $\{u, v_1\}$ is a \emph{bridge} in $G$) and let $S$ denote the set of vertices reachable from $u$ without passing through $v_1$. Since $\deg(v) \geq 2$, there is at least one vertex $w \neq v_{k-1}$ directly connected to it. Notice also that, since $G$ is by assumption in SSE, for any vertex $u' \in S$, we have that $W_{G[S]}(u) \leq W_{G[S]}(u')$ (which is why $v_1$ prefers to connect to $u$ among all other vertices in $S$ \footnote{In particular, the inequality $W_{G[S]}(u) \leq W_{G[S]}(u')$ is a direct consequence of the SSE conditions and the fact that $\{u, v_1\}$ is a bridge. Indeed, for any vertex $u'$ in $S$, let $G'$ be the graph produced by swapping $\{v_1, u\}$ with $\{v_1, u'\}$. Then, by the SSE conditions $W_G(v_1) \leq W_{G'}(v_1)$. Also, since $\{v_1, u\}$ (respectively $\{v_1, u'\}$) is a bridge in $G$ (respectively $G'$), we have that 

\begin{displaymath}
W_G(v_1) = W_{G[S]}(v_1) + W_{G[V-S]}(v_1) = W_{G[S]}(u) +|S| + W_{G[V-S]}(v_1).
\end{displaymath}
An identical expression holds for $W_{G'}(v_1)$, and since $W_{G'[V-S]}(v_1) = W_{G[V-S]}(v_1)$, we can conclude that $W_{G[S]}(u) \leq W_{G[S]}(u')$.}). In particular, this is true for any vertex $z$ at maximum distance from $u$ in $G[S]$. Having vertices $v, u, w$ and $z$ in mind, we now distinguish two cases:

\begin{enumerate}

\item If $|S| \leq \frac{n}{2}$, then $z$ would prefer swapping any of its edges leading to $u$ in order to connect directly to $v_1$ (see also Figure \ref{Lemma1firstcase} for reference). Indeed, if $G'$ is the graph after the swap, then the new cost of $z$ will be 


\begin{eqnarray}
W_{G'}(z) &=& - dist(u, z) |V-S| + \sum_{x \in V-S} dist_G(x, z) + |S|-1 + \sum_{x \in S-\{z\}} dist_G (x, u) \\
& \leq & 2|S| -n +1  \nonumber +\sum_{x \in V-S} dist_G(x, z) + W_{G[S]}(u) \\
& \leq & 2|S| -n +1  \nonumber + \sum_{x \in V-S} dist_G(x, z) + W_{G[S]}(z) \label{eq3} \\&<& W_{G}(z)  
\end{eqnarray}
\normalsize
where in inequality (\ref{eq3}) we used the fact that $W_{G[S]}(u) \leq W_{G[S]}(u')$ for any vertex $u' \in S$. But then we proved that $z$ has a profitable swap, which is a contradiction because $G$ is by assumption in SSE.

\begin{figure}[ht]
\centering
\includegraphics{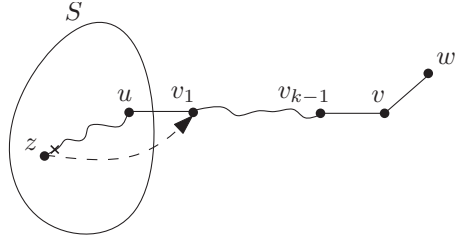}
\caption{An illustration of vertices $u, v, w, z$, the set $S$ and the swap of vertex $z$ to $v_1$.}
\label{Lemma1firstcase}
\end{figure}

\begin{figure}[ht]
\centering
\subfigure[]{
\includegraphics[width=0.2\textwidth]{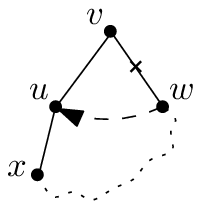}
\label{partB-caseB1}} \qquad \qquad \qquad 
\subfigure[]{
\includegraphics[width=0.2\textwidth]{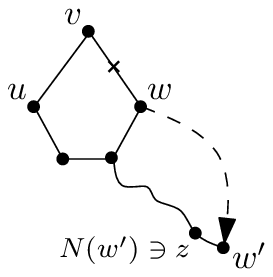}
\label{partB-caseB2}}
\caption{(a) Non-existence of an induced cycle of size strictly greater than 5 including vertices $u, v$ and $w$. (b) Illustration of a profitable swap for $w$ for the case where $(u,w)$ does not exist and $diam(G)\geq 10$.}
\label{fig:R5hopsrate}
\end{figure}

\item The case $|S| \geq \frac{n}{2}$ is similar by symmetry: $v_1$ will have the role of $u$, $w$ will have the role of $z$ (i.e. the vertex at maximum distance from $v_1$ in $V-S$, and $z$ will have the role of $w$ (i.e. a neighbor of $u$ different from $v_1$). Notice then that since $|V-S| \leq \frac{n}{2}$, we can use an identical reasoning with that of the previous case in order to prove that $w$ has a profitable swap (i.e. deleting any of its edges leading to $v_1$ to connect to $u$), which is a contradiction because $G$ is by assumption in SSE.

\end{enumerate}
This shows that there are at least two paths from $u$ to $v$ that the first edge on each path is different. The corresponding proof for $v-u$ paths is similar and so the proof of part (A.) of the lemma is complete.

\emph{Part (B.):} Let $v$ be a vertex of degree 2 and let $u, w$ be its two neighbors. Assume for the sake of contradiction that $diam(G) \geq 10$. Then there is a vertex $w'$ such that $dist(w, w') \geq 5$. We now distinguish two cases: 

\begin{itemize} 

\item[(a)] If the edge $(u, w)$ exists, then it is profitable for $w$ to swap $(w, v)$ with $(w, w')$. Indeed, this swap increases the distance of $w$ from $v$ by 1, decreases its distance from $w'$ by at least 4 and it does not increase its distance from any other vertex. 

\item[(b)] If edge $(u, w)$ does not exist, then any vertex in $N_G(u) \cup N_G(w)$ must belong to a cycle of length at most 5 which includes $v, u$ and $w$ (if there was a vertex $x$ which does not belong to such a small cycle and assuming, without loss of generality, that $x \in N_G(u)$, then we come to a contradiction, because it is profitable for $w$ to swap $(w, v)$ with $(w, u)$ -- for reference, see Figure~\ref{partB-caseB1}; this swap increases the distance of $w$ from $v$ by 1, decreases its distance from $u$ by 1, decreases its distance from $x$ by 1 and it does not increase its distance from any other vertex). But then swapping $(w, v)$ with $(w, w')$ is profitable for $w$. Indeed, this swap increases the distance of $w$ from $v$ by at most 3, increases its distance from $u$ by at most 1, decreases its distance from $w'$ by at least 4, decreases the distance from any neighbor of $w'$ by at least 2 and it does not increase its distance from any other vertex (see Figure~\ref{partB-caseB2}). 

\end{itemize}
So in both cases we have a contradiction to the fact that $G$ is in SSE, which completes the proof of part (B.) of the lemma. \qed

\subsection{A Structural Characterization of Graphs in SSE} \label{structuralcharacterization}

We first give some definitions that will be useful in the statement of Theorem \ref{maintheorem}. For a positive integer $c$ and vertices $u, v \in V$, let $A_{u, v}(c)$ be the set of vertices whose distances from $u$ and $v$ differ by exactly $c$, i.e., setting $k_1 = \min\{dist_G(u, z), dist_G(v, z)\}$ and $k_2 = \max\{dist_G(u, z), dist_G(v, z)\}$, we have

\begin{equation}
A_{u, v}(c) \stackrel{def}{=} \{z \in V \backslash \{u, v\}: k_2 - k_1 = c\}.
\end{equation}

The following Theorem is a strong structural characterization of graphs in SSE.

\begin{theorem} \label{maintheorem}
Let $G = (V, E)$ be a graph in SSE. Then $\sum_{c=0}^{\infty} c |A_{u, v}(c)| \leq \frac{\delta'_{uv}+1}{\delta'_{uv}-1}n$, for any two vertices $u, v \in V$, where $\delta'_{uv} \stackrel{def}{=} \min\{ \deg_{-1}(v), \deg_{-1}(u)\}$.
\end{theorem}
\proof Let $u, v \in V$ be any two vertices at distance at least 2, such that

\begin{equation}
2 \leq \delta'_{uv} \stackrel{def}{=} \min\{ \deg_{-1}(v), \deg_{-1}(u)\}.
\end{equation}

We now perform the following random experiment: With probability $\frac{1}{2}$ pick vertex $u$ and then swap a randomly (and uniformly) chosen edge connecting $u$ with one of its neighbors in $N_{-1}(u)$ with the (previously non-existing) edge $\{u, v\}$. Otherwise, pick vertex $v$ and swap a randomly chosen edge in $N_{-1}(v)$ with the (previously non-existing) edge $\{v, u\}$. We will denote by $G'$ the resulting graph after the swap. Notice that $G$ remains connected after any such swap.


Consider a vertex $z \in V \backslash \{u, v\}$ and define integers $k_1 = \min\{dist_G(u, z), dist_G(v, z)\}$ and also $k_2 = \max\{dist_G(u, z), dist_G(v, z)\}$. Let also $X_z$ denote the random variable of the distance increase between $z$ and the vertex that performed the swap in the above random experiment. Notice that $X_z <0$ if the farthest vertex from $z$ was chosen and $k_2 > k_1 +1$ (see, for example, Figure~\ref{theorem1SSE}). Furthermore, $X_z \geq 0$ if the closest vertex from $z$ was chosen along with an incident edge that happens to be in all shortest paths from the chosen vertex to $z$. Since, there is at most one choice for incident edges in which the latest scenario happens, we conclude that $X_z$ is stochastically dominated by the following random variable:

\begin{figure}[ht]
\centering
\includegraphics[width=0.3\textwidth]{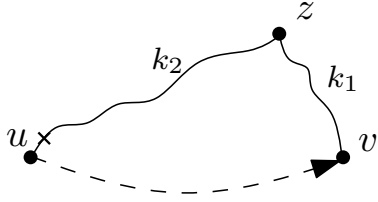}
\caption{A profitable swap happens with probability at least $1/2$.}
\label{theorem1SSE}
\end{figure}

\begin{equation}
Y_z = \left\{
\begin{array}{ll}
	k_1+1 - k_2, & \quad \textrm{with probability $\frac{1}{2}$} \\
	k_2+1 - k_1, & \quad \textrm{with probability $\frac{1}{2} \frac{1}{\delta'_{uv}}$} \\
	0, & \quad \textrm{otherwise.}
\end{array} \right.
\end{equation}
By domination we then have that

\begin{equation} \label{singlevertexcostnew}
\mathbb{E}[X_z] \leq \mathbb{E}[Y_z] = \frac{1}{2 \delta'_{uv}}\left((\delta'_{uv}-1)(k_1 - k_2) + \delta'_{uv}+1 \right).
\end{equation}
Therefore, $\mathbb{E}[X_z]$ is small when $k_1$ is much smaller than $k_2$ and is always at most $\frac{3}{4}$. 

Define now $X$ to be the total increase in the sum of distances of the swapping vertex from every other vertex in $V \backslash \{u, v\}$. By linearity of expectation, equation (\ref{singlevertexcostnew}) and the definition of $A_{u, v}(c)$ we have 

\begin{eqnarray}
\mathbb{E}[X] & = & \sum_{z \in V \backslash \{u, v\}} \mathbb{E}[X_z] \nonumber \\
& \leq & \sum_{c=0}^{\infty} \sum_{z \in A_{u, v}(c)} \frac{1}{2 \delta'_{uv}}\left(\delta'_{uv}+1 - (\delta'_{uv}-1) c \right) \\
& = & \sum_{c=0}^{\infty} \frac{\delta'_{uv}-1}{2 \delta'_{uv}}\left(\frac{\delta'_{uv}+1}{\delta'_{uv}-1} - c \right) |A_{u, v}(c)|.
\end{eqnarray}
The right hand side of the above inequality becomes negative when $\sum_{c=0}^{\infty} c |A_{u, v}(c)| > \frac{\delta'_{uv}+1}{\delta'_{uv}-1}n$. By the probabilistic method $\mathbb{E}[X]<0$ would imply that there is a swap between $u$ and $v$ that improves the cost of the swapping vertex. But this is a contradiction, since $G$ is in SSE. This completes the proof of the Theorem. \qed

We now give an alternative, useful interpretation of Theorem \ref{maintheorem}. For any two vertices $u, v$ and a randomly chosen vertex $Z \in V - \{u, v\}$, let $D_{u, v}(Z) = |dist(u, Z) - dist(v, Z)|$. Namely, $D_{u, v}(Z)$ is the random variable of the absolute difference of the distance of $Z$ from $u, v$. We then have the following:

\begin{corollary}
Let $G = (V, E)$ be a graph in SSE. Then, for any two vertices $u, v$ of degree greater than 1, we have that $\mathbb{E}[D_{u, v}(Z)] \leq 3$.
\end{corollary}   
Notice that the constant 3 in the above Theorem can be made smaller if we have additional information of the degrees of each vertex. For example, provided $G$ is in SSE and has no vertex of degree 1, then by the second part of Lemma \ref{SSEgraphstructure} we have the following: Either $diam(G) \leq 9$, or $\mathbb{E}[D_{u, v}(Z)] \leq \frac{4}{3}$, for any two vertices $u, v \in V$.


\subsection{Large $k$-Vicinity and Dense Graphs} \label{largeanddense}

Using the structural characterization of Theorem \ref{maintheorem} we can prove bounds on the diameter of SSE graphs that are either dense or have large maximum degree. The following result relates the size of the largest $k$-vicinity of a graph $G$ in SSE to its diameter $diam(G)$.

\begin{theorem} \label{maxdegreecase}
Let $G = (V, E)$ be a graph in SSE and let $\Delta^{(k)} = \max_u |B_u(k)|$. Then $diam(G) \leq \frac{6n}{\Delta^{(k)}}+2+4k$, for any $k>0$.
\end{theorem}
\proof Let $u \in V$ be such that $|B_u(k)| = \Delta^{(k)}$ and let $u'$ be a vertex with $\deg(u') \geq 2$ which is at maximum distance from $u$. Then $dist(u, u') \geq \frac{diam(G)}{2} -1$. Furthermore, by the triangle inequality, for any $w \in B_u(k)$ we have that

\begin{equation}
dist(w, u') \geq dist(u, u') - dist(u, w) \geq \frac{diam(G)}{2} -1 - k.
\end{equation}
Therefore, there are at least $\Delta^{(k)}$ vertices $w$ for which $D_{u, u'}(w) = |dist(w, u') - dist(w, u)| \geq \frac{diam(G)}{2} - 1 - 2k$. Choosing now a vertex $Z$ uniformly at random we have

\footnotesize
\begin{eqnarray}
\mathbb{E}[D_{u, u'}(Z)] & \geq & \frac{1}{n} \left( \sum_{z \in N(u)} D_{u, u'}(Z) + \sum_{z \notin N(u)} D_{u, u'}(Z) \right) \\
& \geq & \frac{\Delta^{(k)}}{n} \left(\frac{diam(G)}{2} - 1 - 2k \right).
\end{eqnarray}
\normalsize
Applying Theorem \ref{maintheorem} we must have that $\frac{\Delta^{(k)}}{n} \left(\frac{diam(G)}{2} - 1 - 2k \right) \leq 3$, which completes the proof. \qed

Notice that, by Theorem \ref{maxdegreecase}, if (say) for $k = \frac{diam(G)}{5}$ we have $\Delta^{(k)} = \frac{n}{poly(\log{n})}$, then the diameter of the SSE graph is polylogarithmic. Furthermore, since for $k=1$ we have that $|B_u(1)| = \deg(u)+1$, we can conclude the following easy Corollary concerning the diameter of large degree graphs in SSE:

\begin{corollary} \label{maxdegreecorollary}
Let $G = (V, E)$ be a graph in SSE and let its maximum degree be such that $\Delta \geq \frac{n}{\log^l{n}}$, for some $l>0$. Then $diam(G) = O(\log^l{n})$.
\end{corollary}
We also note that the upper bounds on the diameter of a graph in SSE that we can prove using directly either Lemma 10 or Corollary 11 of \cite{ADHL10} are weaker than the ones we proved here using Theorem \ref{maintheorem} by a factor of $\Theta(\log{n})$.

We can also prove the following result relating the diameter of a graph $G$ in SSE to the number of its edges.

\begin{theorem} \label{densecase}
Let $G = (V, E)$ be a graph of minimum degree at least 2 and let $e(G)$ be the number of its edges. If $G$ is in SSE, then $diam(G) \leq \frac{6n^2}{e(G)+ \frac{n}{2}} +4$. In particular, if $e(G) \geq \frac{n^2}{\log^l{n}}$, for some $l>0$, then $diam(G) \leq O(\log^l{n})$.
\end{theorem}
\proof Set $k_0 = \frac{diam(G)}{8}$. By Theorem \ref{maxdegreecase}, for any $u \in V$ we have that 

\begin{equation}
B_u(k_0) \leq \frac{12n}{diam(G) - 4}.
\end{equation}
This implies that there are at least $\frac{1}{2}n \left(n-\frac{12n}{diam(G) - 4} \right)$ pairs of vertices $u, v$ which have $dist(u, v) \geq k_0+1$. The number of such pairs of vertices must be at most the number of non-edges in $G$. Therefore,

\begin{equation}
{n \choose 2} - e(G) \geq \frac{1}{2}n \left(n-\frac{12n}{diam(G) - 4} \right)
\end{equation} 
from which we get the desired bound. \qed

\section{A Model of Local Costs} \label{ourmodel}

We now define our \emph{Local Cost Network Creation Game}, which is simpler than the model of \cite{ADHL10} and also admits an exact potential. Let $G = (V, E)$ be any undirected graph with $n$ nodes. As in the model of \cite{ADHL10}, the players in our game can be identified as the set of vertices of the graph, and any player $u \in V$ can swap one of its incident edges (which defines the set of available actions for each player). 
In contrast to \cite{ADHL10} however, the payoff of a vertex depends only on the structure of its immediate neighborhood and not on the entire network. In particular, we define the \emph{profit} of $u \in V$ in $G$ as $\gamma_G(u) = \sum_{v \in N_G(u)} \deg_G(v)$, i.e., the profit of $u$ is the sum of the degrees of its neighbors\footnote{A natural generalization is to consider nodes at distance at most $k$ from $u$.}. A \emph{profitable swap} is an edge swap that improves (increases) the profit of the vertex that performs it. 

Notice that an arbitrary sequence of profitable swaps (by nodes $v_1, v_2, \ldots$) actually transforms the initial graph through a sequence of \emph{configuration graphs} $G_0, G_1, G_2, \ldots$. We will write $G_i \stackrel{v_i}{\rightarrow} G_{i+1}$ and mean that configuration $G_i$ produces configuration $G_{i+1}$ by a selfish swap by vertex $v_i$. Vertex $v_i$ is called \emph{deviator} in configuration $G_i$. A graph $G$ is a \emph{local cost swap equilibrium configuration} if no vertex can perform a selfish (improving) swap. We note the following:

\begin{theorem}
If $G$ is a local cost swap equilibrium configuration, then it contains a star as a spanning subgraph.
\end{theorem}
\proof Notice that a vertex $u$ does not have a profitable swap in $G$ when $\deg_G(v) > \deg_G(w)$, for any $v \in N_G(u)$ and $w \notin N_G(u)$. This means that $u$ connects to all vertices of maximum degree. Moreover, either all vertices are connected to all vertices of maximum degree and the graph contains a star as a spanning subgraph, or there is some vertex $u$ not connected to at least one vertex $w$ of maximum degree, in which case $u$ can benefit from swapping one of its edges to connect to $w$ and thus increase the maximum degree. \qed


\subsection{An Exact Potential} \label{potentialproof}

We now show that our Local Cost Network Creation Game admits an exact potential function.

\begin{theorem} \label{potential}
The function $\Phi(G) = \frac{1}{2} \sum_{v \in V(G)} \deg_G(v)^2$ is an exact potential for the Local Cost Network Creation Game.
\end{theorem}
\proof Consider a profitable swap performed by vertex $u$, which swaps edge $\{u, v\} \in E(G)$ with edge $\{u, w\} \notin E(G)$ and let $G'$ be the resulting graph. Then the following are true: (a) $\deg_{G'}(u) = \deg_{G}(u)$, (b) $\deg_{G'}(v) = \deg_{G}(v)-1$, (c) $\deg_{G'}(w) = \deg_{G}(w)+1$ and (d) the degree of any other vertex remains unchanged. Therefore

\begin{eqnarray} 
\gamma_{G'}(u) - \gamma_G(u) & = & \sum_{z \in N_{G'}(u)} \deg_{G'}(z) - \sum_{z \in N_G(u)} \deg_G(z) \\
& = & \deg_G(w)+1 -\deg_G(v).
\end{eqnarray}
The corresponding change in the value of the function $\Phi(\cdot)$ is then

\begin{eqnarray}
\Phi(G') - \Phi(G)  & = & \frac{1}{2} \left( \deg_{G'}(v)^2 + \deg_{G'}(w)^2\right) - \frac{1}{2}\left(\deg_G(v)^2 - \deg_G(w)^2 \right) \\
& = & \deg_G(w) - \deg_G(v) +1 \\
& = & \gamma_{G'}(u) - \gamma_G(u)
\end{eqnarray}
which proves that $\Phi(\cdot)$ is an exact potential for our game. \qed

\subsection{Reaching Equilibrium using a Limited Number of Queries} \label{lowqueries}

By Theorem \ref{potential}, an equilibrium graph can be found in at most $O(n^3)$ time steps (egde swaps), starting from any initial graph. However, in order for a vertex $u$ to compute a better response (i.e., a profitable swap), it requires information about the degree from all non-adjacent vertices in the graph, i.e., all $v \in V \setminus N_G(u)$. In many cases though, especially for large-scale networks like the Internet, it is inefficient to acquire such information about all the nodes in the network. On the other hand, we can assume that any vertex $u$ can get such information for a limited (e.g., constant) number of non-neighboring nodes by asking an oracle (this setup is also common in the literature of property testing in graphs, see for example \cite{F01}). In this setup the following holds:

\begin{theorem} \label{limitedqueriestheorem}
If any vertex $u$ can obtain information about the degree of $c \geq 1$ randomly chosen non-neighboring vertices, then our network formation game can converge in an equilibrium graph in a polynomial expected number of steps.  
\end{theorem}
\proof Let $G_0 = (V_0, E_0)$ be the current graph. Consider the following procedure: At any time $t \geq 1$, select a vertex $u$ uniformly at random from $V$ and then ask the oracle to reveal the degrees of $c$ randomly chosen non-neighbors of $u$ (we assume that $u$ knows the degrees of its neighbors), namely $v_1, \ldots v_c \in V_t - N_{G_t}(u)$. If one of the vertices among $v_1, \ldots, v_c$ has degree equal or larger than the degree of some neighboring vertex of $u$ in $G_t$, then $u$ performs a profitable swap. Otherwise, it does nothing. The resulting graph will be denoted by $G_t= (V_t, E_t)$.

Notice now that, if at any time step $t$, $G_t$ is not an equilibrium graph, then by Theorem \ref{potential} there must be at least one profitable swap. In particular, there exist vertices $u, v, w$ such that $w \in N_{G_t}(u)$, $v \notin N_{G_t}(u)$ and $\deg_{G_t}(w) \leq \deg_{G_t}(v)$. The probability that $u$ is selected and also $v$ is among the randomly chosen non-neighbors of $u$ is at least $\frac{c}{n^2}$. 

Consider now the stochastic process $\{X_t\}_{t \geq 0}$, where $X_t = \frac{1}{2} \sum_{z \in V_t} \deg_{G_t}(z)^2$. Notice then that, provided $G_t$ is not an equilibrium graph, $\Pr(X_{t+1} \geq X_t +1) \geq \frac{c}{n^2}$ and also $\Pr(X_{t+1} = X_t) = 1 - \Pr(X_{t+1} \geq X_t +1)$. Notice also that the absorbing states of the stochastic process $\{X_t\}_{t \geq 0}$ correspond to equilibrium graphs, and we have $0 \leq X_t \leq \frac{n^3}{2}$ for any $t$ and in particular, for any equilibrium graph $G_t$. 

From the above, we conclude that the number of steps needed for $\{X_t\}_{t \geq 0}$ to reach an absorbing state is stochastically dominated by a geometrically distributed random variable $Geom\left(\frac{n^3}{2}, \frac{c}{n^2} \right)$. Therefore, the mean number of steps needed for absorption is at most $\frac{n^5}{2c}$, which completes the proof. \qed

We note also that we can decide whether the procedure in the proof of Theorem \ref{limitedqueriestheorem} has reached an equilibrium graph with high probability. Indeed, if after at least $\Omega(n^3)$ steps no swap has occurred, then by the Markov inequality, we can correctly (positively) decide whether we have reached equilibrium with probability at least $1 - O\left( \frac{1}{n}\right)$.

\section{Conclusions and Future Work} \label{conclusions}

In this paper we considered network formation games based on the swap operation. In particular, for the sum version of the model of Alon et al. \cite{ADHL10} we provided a new, structural characterization of equilibrium graphs (Theorem \ref{maintheorem}), according to which, for any two vertices of degree greater than 1, the majority of the rest of the vertices are almost equidistant from them. By a direct application of Theorem \ref{maintheorem} we could prove poly-logarithmic upper bounds on the diameter of SSE graphs that are either dense enough or have large $k$-vicinity (thus partially settling positively a conjecture of \cite{ADHL10} for these cases). It remains open whether we can use the full power of Theorem \ref{maintheorem} to provide more general and stronger bounds on the diameter of graphs in SSE. 

As a second contribution, we defined in this paper a new network formation game, which is also based on the swap operation, but the cost for each player/vertex depends only on the degrees of its neighbors. We proved that this network creation game admits an exact potential, and also that any equilibrium graph contains an induced star. Furthermore, we considered a case where vertices can only acquire limited knowledge concerning non-neighboring vertices and we showed that, even in this case, we can reach equilibrium in expected polynomial time. Providing bounds for the price of anarchy in this model is left as an open problem for future research. Finally, we intend to study extensions of our model of local costs, in which the profit for each vertex depends on the structure of its $k$-vicinity.

%
%

\end{document}